\documentclass{PoS}

\usepackage[utf8]{inputenc}
\usepackage{graphicx}
\usepackage{subfigure}
\usepackage{amsmath,amssymb}

\def\Tr{{\rm Tr}\,}

\title{Large-$N$ reduction with adjoint Wilson fermions}
\ShortTitle{Large-$N$ reduction with adjoint Wilson fermions}

\author{Barak Bringoltz,$^{ab}$ \speaker{Mateusz Koren}$^{cb}$ and Stephen
R.~Sharpe$^{b}$\\
\llap{$^a$}IIAR -- the Israeli Institute for Advanced Research,\\
Rehovot, Israel\\
\llap{$^b$}Department of Physics, University of Washington,\\
Seattle, WA 98195-1560, USA\\
\llap{$^c$}M. Smoluchowski Institute of Physics, Jagiellonian University,\\
Reymonta 4, 30-059 Cracow, Poland\\
E-mail: \email{barak.bringoltz@gmail.com},
\email{mateusz.koren@uj.edu.pl}, \email{sharpe@phys.washington.edu}}

\abstract{We analyze the large-$N$ behavior of $SU(N)$ lattice gauge theories with adjoint fermions by studying
volume-reduced models, as pioneered by Eguchi and Kawai. We perform simulations on a single-site lattice for $N_f=1$ and
$N_f=2$ Wilson Dirac fermions with values of $N\leq53$. We show for both values of $N_f$ that in the large-$N$ limit
there is a finite region, containing both light and heavy fermions, of unbroken center symmetry where the theory
exhibits volume independence. Using large-$N$ reduction we attempt to calculate physical quantities such as the string
tension and meson masses.}
        
\FullConference{The 30 International Symposium on Lattice Field Theory\\
Cairns, Australia\\
June 24-29, 2012}

\begin{document}

\section{Introduction}

The idea of volume independence of large-$N$ gauge theories dates back to the paper by Eguchi and Kawai \cite{EK82}.
They have shown that, provided certain conditions hold, in the large-$N$ limit the pure gauge $SU(N)$ theory defined
on a single-site lattice\footnote{To be precise Eguchi and Kawai discussed the $U(N)$ gauge theory that however
coincides with $SU(N)$ as $N\to\infty$.} (the so-called Eguchi-Kawai model) has the same Wilson-loop expectation values
as the theory defined on arbitrary (including infinite) volume.

The only non-trivial assumption of their prescription is the lack of spontaneous breaking of the $\mathbb{Z}_N$ center
symmetry. This assumption was quickly shown to be false \cite{BHN82}. The center symmetry in pure gauge theory is
spontaneously broken below some critical lattice size $L_\text{crit}$ as can be seen both by lattice calculations
\cite{BHN82,OK82,KNN03} and in perturbation theory \cite{BHN82,KM82}.

Several ways to fix Eguchi-Kawai reduction were proposed over the years. Let us quickly sketch the three particularly
interesting approaches:
\begin{enumerate}
\item{Use twisted boundary conditions (the so-called Twisted Eguchi-Kawai or TEK model) \cite{TEK1,TEK2}. 
While the original choice of twists has been shown not to work~\cite{Bietenholz:2006cz,TV07,Azeyanagi:2007su}
a different choice appears to overcome the problems~\cite{TEK3,TEK4}.}
\item{Always work with $L > L_\text{crit}(b)$ \cite{KNN03}. This idea known as partial reduction or continuum
reduction\footnote{See also Refs. \cite{NN03,KOR09} for analysis in three dimensions.}, allows one to make simulations
directly in the pure gauge theory at the price of having to work with a box of finite physical size (thus with
$L_\text{crit}\to\infty$ in the continuum limit).}
\item{Use adjoint fermions to stabilize the center symmetry. This idea, inspired by the orbifold large-$N$ equivalences
\cite{HN02,KUY04}, was proposed in Ref.~\cite{KUY07} and is the basis of our analysis\footnote{There is also a
related idea of trace-deformed reduction \cite{UY08,HV11} that however becomes rather complex when reducing more than
one dimension.}.}
\end{enumerate}

In this paper we discuss the single-site lattice model with adjoint fermions -- the Adjoint Eguchi-Kawai (AEK) model.
It is shown in Ref.~\cite{KUY07} that by adding $N_f>1/2$ massless adjoint Dirac fermions obeying periodic
boundary conditions, the center symmetry remains intact at the one-loop level. Perturbatively, fermions with masses larger
than $\sim 1/(aN)$ were expected to break the center symmetry~\cite{HM09,BB09}. However several 
lattice calculations~\cite{BS09,Catterall,Hietanen} 
found that the center symmetry is likely to be intact for rather heavy fermions of
mass $\sim 1/a$, both for $N_f=1$ and $N_f=2$. In Ref.~\cite{BKS11} we argued that this result holds in the large-$N$
limit for $N_f=2$. In this paper we present evidence that this is also most likely the case for $N_f=1$. This result was
given a semi-analytic understanding (using arguments going beyond perturbation theory) in Ref.~\cite{AHUY}.

Our results are of considerable phenomenological interest. Volume reduction in the region of heavy fermions,
where the dynamics is governed mostly by gauge degrees of freedom, is a realization of the original idea of Eguchi and Kawai.
Also, the $N_f=1$ AEK model can be connected by a chain of large-$N$ equivalences to the Corrigan-Ramond large-$N$ limit of
two-flavour QCD \cite{KUY07,BS09}. The $N_f=2$ AEK model, on the other hand,
 is supposed to behave similarly (e.g. with respect to
the $\beta$-function) as $SU(2)$ gauge theory with 2 adjoint flavours which is most likely a conformal field theory (see
Ref.~\cite{DSS11} and references therein).

Motivated by the success of the reduction we attempt to extract large-distance quantities such as string tension and
meson masses from our simulation. We discuss the $N$ scaling of the effective volume
and its impact on practicality of calculations in reduced models.

\section{Range of quark mass where volume reduction holds}

We analyze the single-site $SU(N)$ lattice model with adjoint Wilson fermions by means of Monte Carlo simulations. The
partition function is
\begin{equation}
	\mathcal{Z}=\int D[U,\psi,\bar\psi]\,e^{(S_{\rm gauge} +
  \sum_{j=1}^{N_f} \bar \psi_j \, D_{\rm W} \, \psi_j)},
\end{equation}
where $S_{\rm gauge}$ is the single-site equivalent of the Wilson plaquette action:
\begin{equation}
S_{\rm gauge} = 2 N b\, \sum_{\mu<\nu} {\rm Re}\Tr U_\mu U_\nu U^\dagger_\mu
U^\dagger_\nu
\end{equation}
($b=\tfrac{1}{g^2 N}$ being the inverse 't Hooft coupling) and $D_W$ is the Wilson Dirac operator on a single site (with
periodic boundary conditions in every direction):
\begin{equation}
D_W = 1 - \kappa \sum_{\mu=1}^4 \left[ \left( 1 - \gamma_\mu\right) U^{\rm
adj}_\mu + \left(1 + \gamma_\mu\right)U^{\dag {\rm adj}}_{\mu}\right],
\end{equation}
where the hopping parameter $\kappa$ is related to the bare fermion mass by
\begin{equation}
m_0 = \frac1{2\kappa}-4.
\end{equation}
Note that the critical value $\kappa_c$ (where the fermion mass vanishes) is shifted from $1/8$ to a larger value due to
additive renormalization and only becomes $1/8$ in the continuum limit.
 
Our numerical simulations are performed using the Hybrid Monte Carlo algorithm (equipped with the rational approximation
in the $N_f=1$ case). We have performed extensive scans of the $\kappa-b$ plane to find the phase structure of the model
(see Ref.~\cite{BKS11} for details on the $N_f=2$ case) reaching $N\leq53$. Our main observable for the detection of
breaking of the center symmetry are the ``open loops'':
\begin{equation}
K_{n}\equiv \frac1{N}\Tr\, U^{n_1}_1\, U^{n_2}_2\, U^{n_3}_3\, U^{n_4}_4, 
\quad {\rm with}\ \ n_\mu =0,\pm 1, \pm 2, \dots
\end{equation}
where $U^{-n}=U^{\dagger n}$. Those results were supplemented by histograms of phases of link eigenvalues.

The results for $N_f=1$ and $N_f=2$ are qualitatively similar. In both cases we find that, at fixed $b$, when decreasing
the fermion mass from infinity ($\kappa=0$) we go from a phase with completely broken center symmetry (as in the pure
gauge model) to phases with partial breakings of $\mathbb{Z}^4_N$. When $\kappa$ exceeds a critical value $\kappa_f(b)$
(or, equivalently, when $m$ is smaller than $m_f$) we enter the so-called ``funnel'' where the center symmetry is
unbroken and volume reduction holds.

\begin{figure}[tbp!]
\subfigure[$N_f=2$] {\label{fig:funnel2} \includegraphics[width=7.25cm]{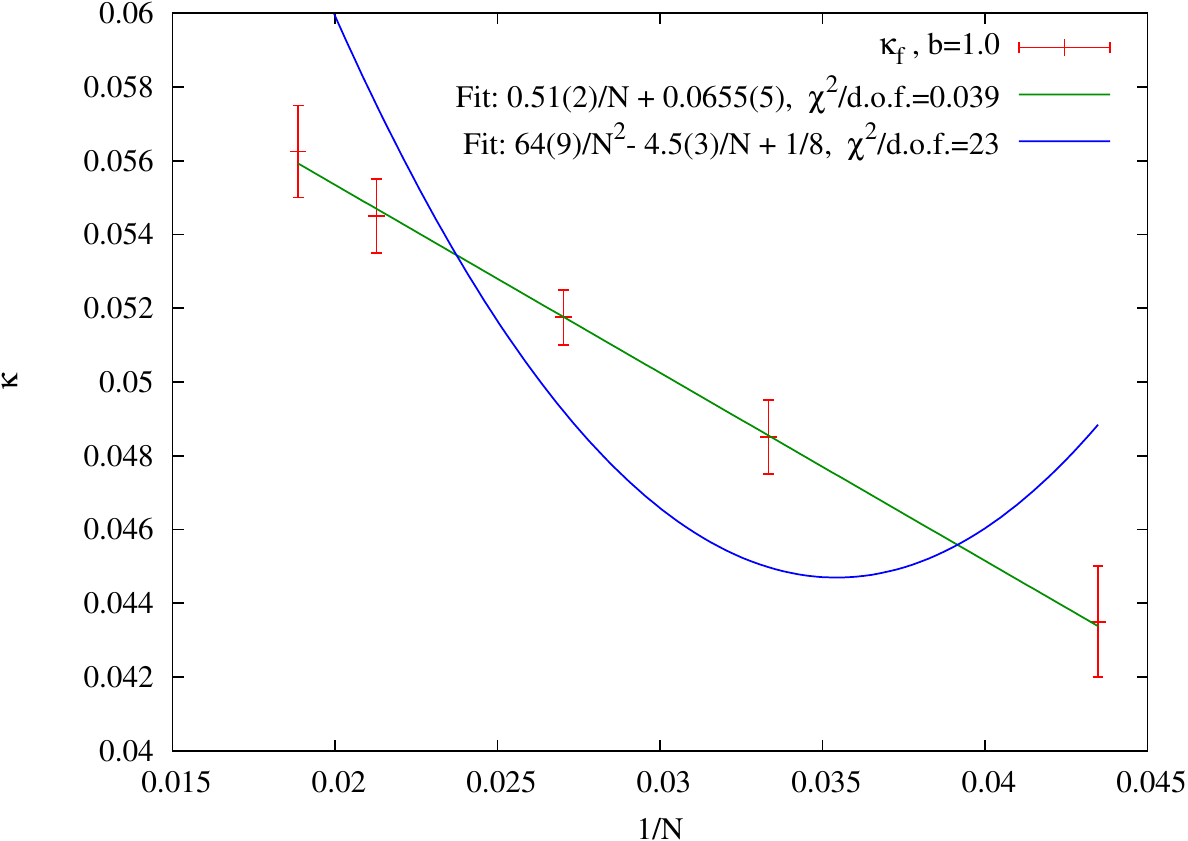}}
\hspace{0.25cm}
\subfigure[$N_f=1$] {\label{fig:funnel1} \includegraphics[width=7.25cm]{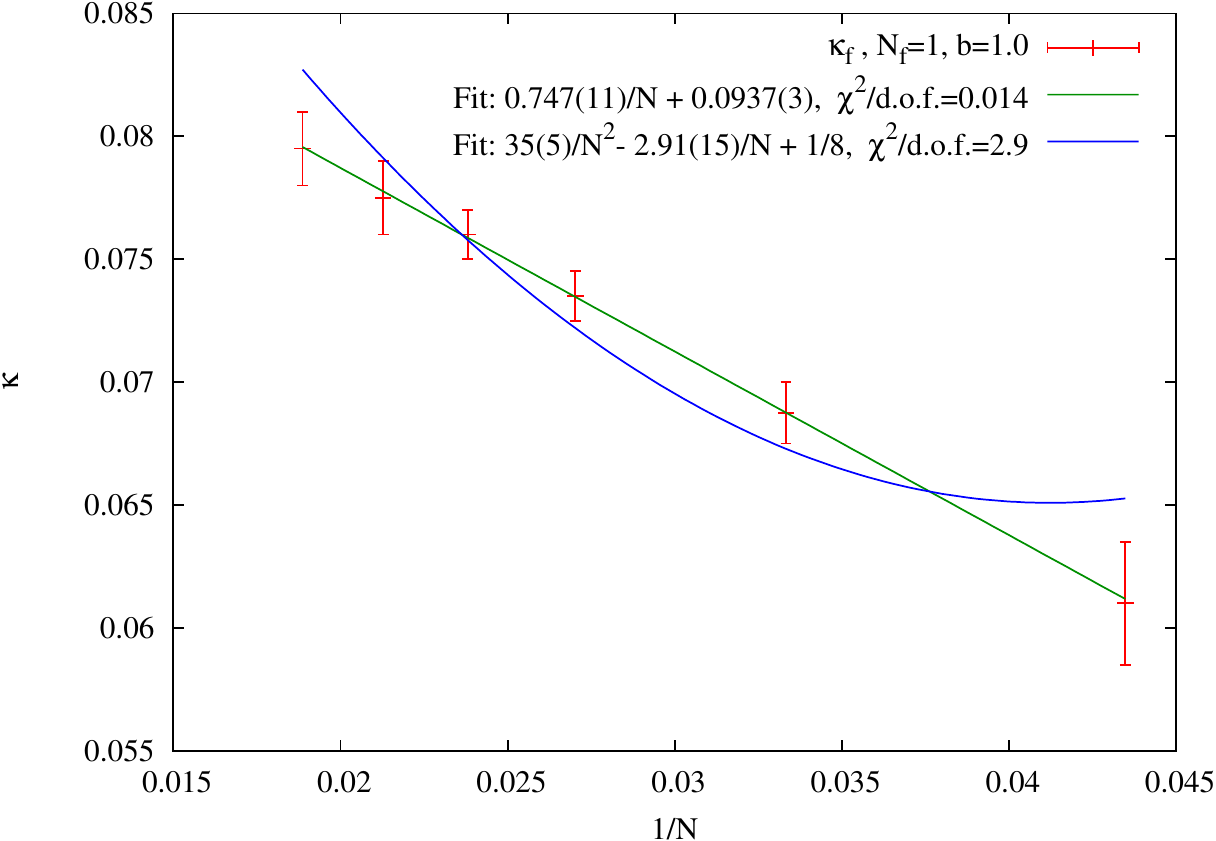}}
\caption{Large-$N$ extrapolations of $\kappa_f$. Note that the errors are rather conservative as the main goal was the
exclusion of the ``closed funnel'' hypothesis.}
\label{fig:funnel}
\end{figure}

An important question is whether the funnel remains of finite width in lattice units as $N\to\infty$.
In Ref.~\cite{BKS11} we performed an extrapolation to the large-$N$ limit at $b=1$ and found
\begin{equation}
\kappa_f(b=1,\,N=\infty) = 0.0655(5),
\end{equation}
as shown in Fig.~\ref{fig:funnel2}. 
Since we know that $\kappa_c> 0.125$,
this implies that the edge of the funnel lies below $\kappa_c$ even at $N=\infty$.
If we attempt to force a fit with $\kappa_f\to 0.125$ when $N\to \infty$, we find
$\chi^2/\text{d.o.f.}=23$, as shown in the figure.
We have now performed a similar calculation in the $N_f=1$ case, obtaining
\begin{equation}
\kappa_f(b=1,\,N=\infty)=0.0937(3),
\end{equation}
as shown in Fig.~\ref{fig:funnel1}.
Thus the funnel is narrower at $N_f=1$.
The fit assuming $\kappa_f\to 0.125$ at $N=\infty$ now has $\chi^2/\text{d.o.f.}=2.9$,
so, while the closure of the funnel is not definitely
excluded, we view this possibility as unlikely 
(particularly in light of the fact that $\kappa_c>0.125$ due to additive
renormalization, making the fit poorer).

\section{Physical quantities: preliminary results \& outlook}

In Ref.~\cite{BKS11} we showed that the AEK model with heavy quarks reproduces the large-volume pure gauge values of the
plaquette very well. The natural next step using volume reduction is to calculate some long-range quantity such as the
string tension. This can be done by calculating the static quark potential from reduced Wilson loops (wrapped many times
around the single-site lattice). However, at our values of $N$ we found it impossible to calculate the quark potential
at separations large enough to give a reliable estimate. This was primarily caused due to large $1/N$ corrections, and
was not an issue of statistical errors.

An example of the observed unphysical behaviour caused by the $1/N$ corrections is the slow rise of Wilson loops at
large separations. In Ref.~\cite{BKS11} we showed that this can be understood qualitatively by inspecting the strong
coupling expansion. On the single site every link in the Wilson loop comes in a pair with its hermitian conjugate 
and thus the zeroth order contribution in the strong-coupling expansion does not vanish.

\begin{figure}[tbp!]
\subfigure[All Wilson loops] {\label{fig:wloops1} \includegraphics[width=7.25cm]{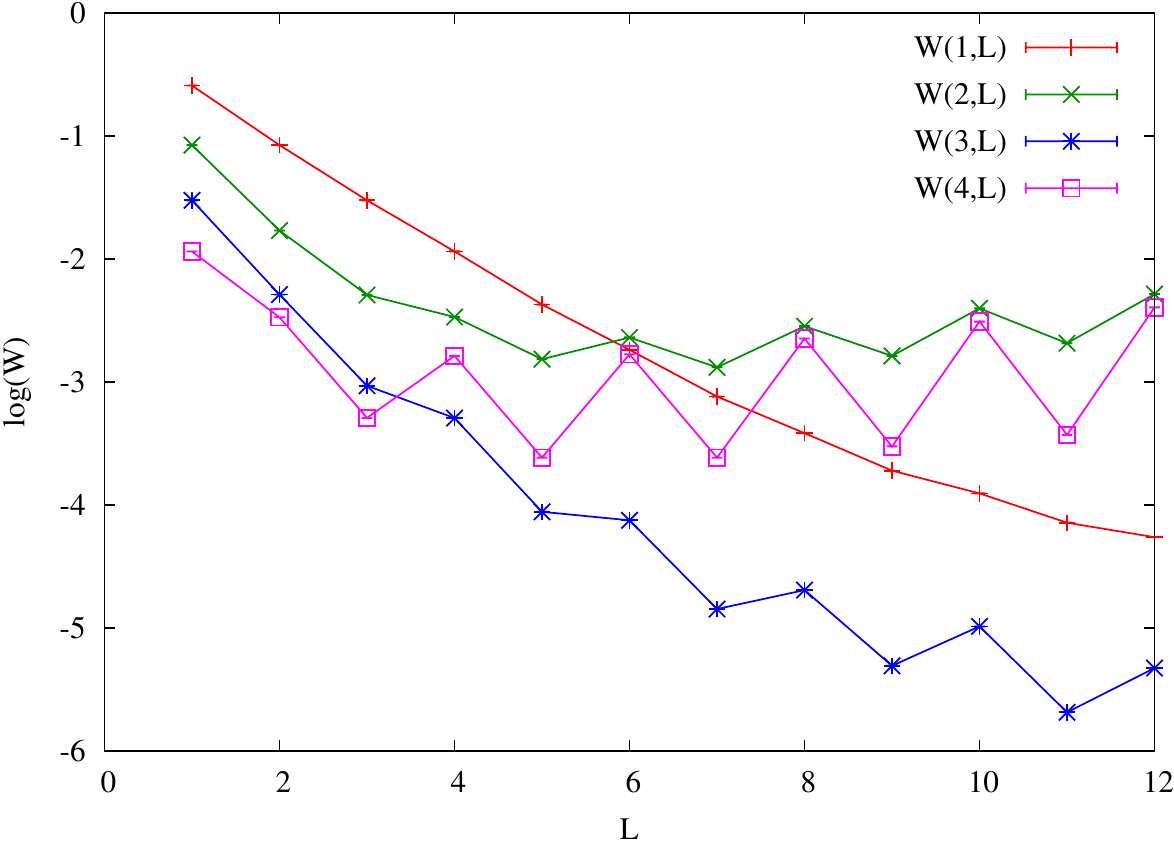}}
\hspace{0.25cm}
\subfigure[Only odd-sized Wilson loops] {\label{fig:wloops2} \includegraphics[width=7.25cm]{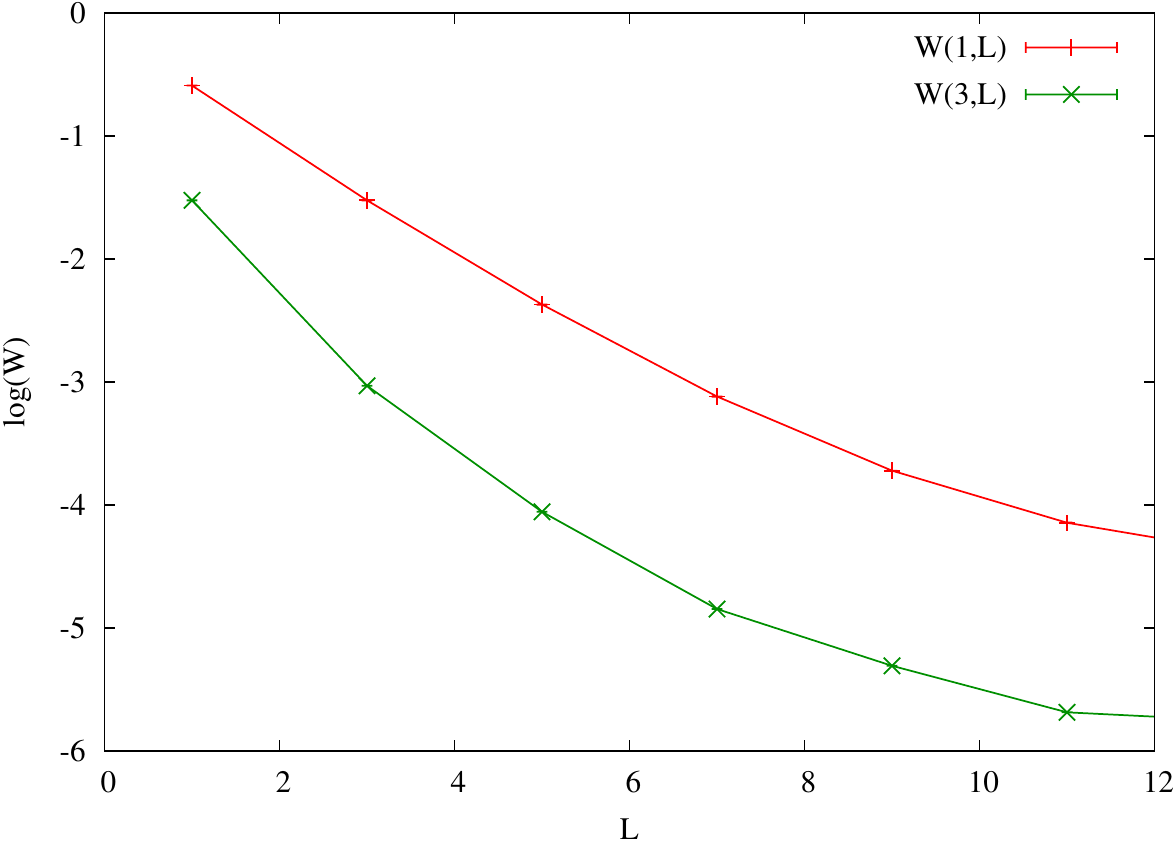}}
\caption{Wilson loops on $2^4$ lattice. $N=10$, $b=0.35$, $\kappa=0.1$.}
\label{fig:wloops}
\end{figure}

One way to tackle this problem is to use $2^4$ lattices. The zeroth order does disappear for the odd-sized loops. In
Fig.~\ref{fig:wloops} we see that indeed the size of finite-$N$ corrections is substantially improved. However, our
values of $N$ to date ($N\le 15$) are too small to allow extraction of the string tension.

Another calculation that one can perform on a single site is finding meson masses using the ``quenched momentum
prescription'' \cite{NN05}. Our first attempt used fundamental Wilson and overlap valence quarks over the sea of heavy
adjoint Wilson fermions -- this can be thought of as a quenched approximation as we expect the heavy adjoint fermions to
have a negligible influence on the propagators. The method of quenched momentum prescription is based on inserting
additional $U(1)$ factors into the temporal gauge links, different for the quark and antiquark. These can be interpreted
as momenta. For the pion the propagator one has
\begin{equation}
\mathcal{M}_\pi(p,m) = \Tr\left\{\gamma_5\,D^{-1}(U_4\,e^{ip/2},m)
  \, \gamma_5\,D^{-1}(U_4\,e^{-ip/2},m) \right\}.
\end{equation}
One can then extract the mass of the pion directly in momentum space. Our first results show sensible behaviour of the
pion mass as a function of $\kappa$, however we also see rather large $1/N$ corrections and the values of $N$ we used
are too small for a reliable large-$N$ extrapolation. 

The beauty of reduction is that both the physical volume and the color degrees of freedom of the unreduced theory are
packaged into the color degrees of freedom of the reduced theory. Thus it makes sense, at least qualitatively, to
introduce an effective box size, $L_{\rm eff}(N)$. (For further discussion of the interpretation of this quantity see
Ref.~\cite{BKS11}.) An important practical issue for numerical simulations is how $L_{\rm eff}$ scales with $N$. The
reasonable possibilities appear bracketed by two cases---$L_\text{eff}\sim N^{1/4}$, as motivated by the orbifold
equivalence, or $L_\text{eff}\sim N^{1/2}$ (as for the Twisted EK model)\cite{AMNS,U04}. (The most optimistic case
$L_\text{eff}\sim N^{1}$ seems to be excluded by the data~\cite{BKS11}). In Ref.~\cite{BKS11} we argued that the Dirac
operator spectrum prefers $L_\text{eff}\sim N^{1/2}$, i.e. the more optimistic scaling behaviour. The situation is,
however, far from clear, with different observables giving different indications. 
We hope that comparisons with results from a $2^4$ lattice may help elucidate the issue.

Another matter is the size of the finite-$N$ corrections on the observables. We clearly see $\mathcal{O}(1/N)$
corrections in the plaquette and the Polyakov loops \cite{BKS11} that are rather large. It it thus promising to use
twisted boundary conditions (the so-called Twisted AEK model~\cite{AHUY, GAO12}) which have only $\mathcal{O}(1/N^2)$
corrections and automatically have $L_\text{eff}\sim N^{1/2}$ scaling.

Finally, it would also be very interesting to compare the meson masses obtained using quenched momentum prescription with a
more standard calculation involving a single elongated direction. Such a setup also allows calculating the glueball
masses.

\section*{Acknowledgments}

This work was supported in part by the U.S. DOE Grant No. DE-FG02-96ER40956, and by the Foundation for Polish Science
MPD Programme co-financed by the European Regional Development Fund, agreement no. MPD/2009/6.

\end{document}